\documentclass[twocolumn,aps,prl,groupedaddress, nobalancelastpage]{revtex4}
\usepackage{graphicx}
\usepackage{color}
\usepackage{amsmath}
\usepackage{marvosym}
\usepackage{bm}
\usepackage{verbatim}

\usepackage{amssymb}
\usepackage{amsthm}
\usepackage{amsfonts}

\newcommand{\be}{\begin{equation}}
\newcommand{\ee}{\end{equation}}
\newcommand{\bea}{\begin{eqnarray}}
\newcommand{\eea}{\end{eqnarray}}

\newcommand{\p}{\partial}

\newcommand{\la}{\langle}
\newcommand{\ra}{\rangle}

\newcommand{\bra}[1]{\la #1|}
\newcommand{\ket}[1]{| #1\ra}

\renewcommand{\vec}[1]{{\bf #1}}
\def\nn{\nonumber\\}

\begin{document}
\title{Cooperative orbital moments and edge magnetoresistance in monolayer WTe$_2$}
\author{Arpit Arora$^1$}
\author{Li-kun Shi$^1$}
\author{Justin C. W. Song$^{1,2}$}
\email{justinsong@ntu.edu.sg}
\affiliation{$^1$Division of Physics and Applied Physics, Nanyang Technological University, Singapore 637371}
\affiliation{$^2$Institute of High Performance Computing, Agency for Science, Technology, and Research, Singapore 138632}

\begin{abstract}
We argue that edge electrons in monolayer WTe$_2$ can possess a ``cooperative'' orbital moment (COM) that critically impacts its edge magnetoresistance behavior. Arising from the cooperative action of both Rashba and Ising spin orbit coupling, COM quickly achieves large magnitudes (of order few Bohr magnetons) even for relatively small spin-orbit coupling strengths. As we explain, such large COM magnitudes arise from an unconventional cooperative spin canting of edge spins when Rashba and Ising spin orbit coupling act together. Strikingly, COM can compete with spin moments to produce an unusual anisotropic edge magnetoresistance oriented at an oblique angle. In particular, this competition produces a direction along which  $\vec B$ is ineffective at gapping out the edge spectrum leaving it nearly gapless. As a result, large contrasts in gap sizes manifest as $\vec B$ is rotated granting giant anisotropic magnetoresistance of $0.1-10$ million $\%$ at $10 \, {\rm T}$ and low temperature.  
\end{abstract} 

\maketitle

Quantum spin Hall (QSH) insulators are highly sensitive to magnetic field. Protected by time-reversal symmetry, QSH insulators exhibit robust gapless edge states and edge electrons that do not backscatter \cite{kane_mele, bhz, qsh_sceince, jps_review, shen_book, finite_size, wu2006, buttiker2009edge, hasan_kane, zhang, qsh_tmds}. This protection is readily lifted when a magnetic field, $\vec B$, is applied to open a gap in the edge spectrum to produce large edge magnetoresistance~\cite{jps_review, qsh_sceince, magnetoconductance} even at relatively low fields.  

In systems with only a small topological band inversion $M \sim 10$ meV such as HgTe and InAs/GaSb quantum well QSH platforms \cite{qsh_sceince, jps_review, shen_book, InAs, InAs_2, InAs_3}, out-of-plane magnetic field induced gap can be very large and is dominated by an orbital effect with giant effective (orbital) $g$ factors of 40-50~\cite{jps_review, magnetoconductance, tarasenko}. In contrast, the QSH insulator monolayer WTe$_2$~\cite{qsh_wte2_1, qsh_wte2_david,  imaging_david, qsh_wte2_pablo} while possessing similar low dissipation transport that persist to high temperature, possesses a large topological band inversion $2M = 1{\rm eV}$~\cite{qsh_tmds, qsh_wte2_1} dominating over other energy scales. As such, ordinary orbital edge magnetoresponse is expected to be severely muted, suppressed by factors of several thousand in comparison with their small $M$ counterparts~\cite{jps_review, magnetoconductance, tarasenko}. 

\begin{figure}
\includegraphics[width=\columnwidth]{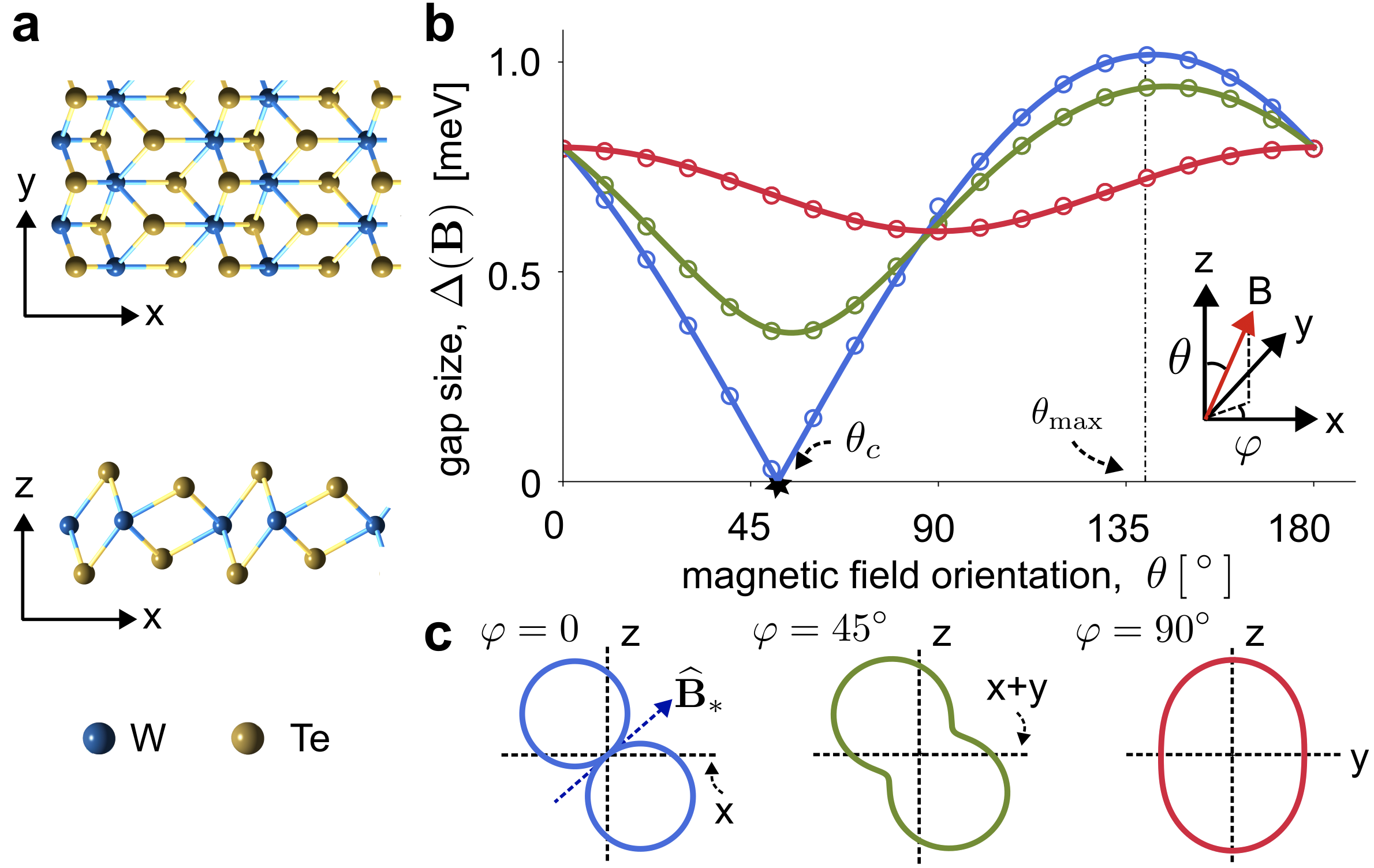}
\caption{(a) Crystal structure of monolayer WTe$_2$ monolayer. (b)  COM yields a significant out-of-plane magnetic field induced gap $\Delta(\vec B = B_z \hat{\vec{z}})$ ($\theta = 0^\circ$) in the edge state spectrum (for an $x$-edge). Away from $\theta = 0^\circ$, $\Delta(\vec B)$ displays an anisotropic angular dependence that depends on azimuthal $\varphi$ and polar $\theta$ angles. 
Here $\textbf{B} = |\textbf{B}|(\cos\varphi\sin\theta, \sin\varphi\sin\theta, \cos\theta)$, and blue, green, and red denote $\varphi = 0, 45^\circ, 90^\circ$ (see inset). 
When $\varphi =0$ (blue), gap nearly vanishes at $\theta_c = 51^\circ$ with a gap value at starred point $\Delta (\vec B_*) = 3.5 \, \mu{\rm eV}$. 
(c) Polar plot of $\Delta(\vec B)$ shown for various planes at $\varphi = 0, 45^\circ, 90^\circ$ (blue, green, red). The starred point in panel (b) manifests in a direction $\widehat{\vec{B}}_*$ (blue dotted arrow) wherein $\vec B$-field is ineffective at gapping the edge spectrum. In all plots, solid lines indicate full gap from Eq.~(\ref{eq:edge}) with effective gyromagnetic coefficients (off-diagonal in Fig.~2; for diagonal, see {\bf SI}). Circles are gaps obtained from edge spectrum ENS, see {\bf SI}. We have taken $\delta_x = 40\, {\rm meV}$, $\delta_z = 70\, {\rm meV}$, and $|\textbf{B}| = 7$ T as an illustration. }\label{fig1}
\end{figure}

Here we argue that an unusual {\it cooperative} effect can buck this expectation to yield a sizable edge orbital response in monolayer WTe$_2$. In particular, we find that the combined action of Rashba spin-orbit coupling (SOC) working together with Ising SOC produces a cooperative orbital magnetic moment (COM). Crucially, COM achieves sizeable values (of order few Bohr magnetons) even when large $M$ far exceeds both Rashba and Ising SOC magnitudes. COM is sustained only when Rashba and Ising SOC coexist. For e.g., in the absence of Ising SOC, COM vanishes, and orbital magnetic response is suppressed by two orders of magnitude. As we explain below, COM is particularly pronounced in WTe$_2$ due to its misaligned Te atoms in the top and bottom layers (see Fig.~\ref{fig1}a) that inextricably link Rashba and Ising SOC~\cite{berry_pablo,likun}. 

COM yields a significant out-of-plane magnetic field induced gap $\Delta (\vec B = B_z \hat{\vec{z}})$ (Fig.~\ref{fig1}b). When $\vec B$ is rotated away from $\hat{\vec z}$, COM competes with spin moments to produce an anisotropic $\Delta(\vec B)$ with a minima canted at an angle oblique to either in-plane or out-of-plane directions (Fig.~\ref{fig1}b,c). Strikingly, $\vec B$-field induced gap almost vanishes at $\theta_c$ when $\vec B$ lies in the $x$-$z$ plane. This defines a direction $\widehat{\vec{B}}_{*} = ({\rm sin} \, \theta_c, 0, {\rm cos}\, \theta_c)$ in three-dimensional space along which magnetic field is ineffective at gapping the edge spectrum even for large magnitudes of $\vec B$; in contrast, when $\vec B$ is directed in other planes (see Fig.~\ref{fig1}b,c), even as the induced gap exhibits anisotropy it remains large along all directions. As we discuss below, the anisotropic gap opening manifests in an anisotropic edge magnetoresistance (AMR), with a giant AMR contrast develops between maximum (at $\theta_{\rm max}$) and minimum (at $\theta_c$) $R(\vec B)$ of order $0.1-10$ million $\%$ at low temperatures and $|\vec B| =  10 \, {\rm T}$.

{\it Gap opening and broken symmetry --} We begin with a symmetry analysis of gap opening in the edge states under an applied magnetic field. For simplicity, we concentrate on QSH edge states along the $x$-edge: 
\be
\label{eq:edge}
H^{\rm edge} (k_x) = \tilde v \sigma_z k_x + \frac{\mu_B}{2}\sum_{ij} g_{ij} \sigma_{i} B_j, 
\ee
where $\sigma_{x,y,z}$ are Pauli matrices that capture a mixed orbital and spin degree of freedom [see Eq.~(\ref{eq:basis})], $\tilde{v}$ is the velocity of the edge state, $i,j = \{x,y,z\}$, and $B_j$ is the magnetic field along $j$ direction. Here $g_{ij}$ are effective gyromagnetic coefficients that link magnetic field applied along $j$ to $\sigma_{i}$ and $\mu_B$ is the Bohr magneton. Here we have taken the lowest symmetry allowed terms, neglecting terms higher order in $k$ and $B$. 

Of particular interest are the off-diagonal terms $g_{xz}$ and $g_{yz}$ that determine gap opening when a $z$-oriented magnetic field is applied; we note that $g_{zz}\sigma_z B_z$ only shifts the edge spectrum leaving it gapless. Symmetry constrains the allowable $B$-field induced terms in Eq.~(\ref{eq:edge}). As an illustration, consider a mirror operation in $z \to -z :\mathcal{M}_z$ (see full discussion in Supplementary Information, {\bf SI}). As expected, $\mathcal{M}_z$ leaves the first term of Eq.~(\ref{eq:edge}) invariant. However, the second term transforms as $\mathcal{M}_{z}^{-1}[g_{ij}\sigma_i B_j] \mathcal{M}_{z}$. As a result, 
when mirror in $z$ is preserved, $g_{xz}$ and $g_{yz}$ terms vanish; 
they are allowed when $\mathcal{M}_z$ is broken. As we will see these arise through Rashba and Ising SOC that mix the spins in the bulk.

{\it Microscopic description of the edge Hamiltonian -- } We now turn to a microscopic description of the edge states. First, we examine a $4\times4$ minimal model for the electronic bulk of monolayer WTe$_2$ \cite{likun,berry_pablo} as $\mathcal{H} = \mathcal{H}_0 + \mathcal{H}_R+ \mathcal{H}_I$. The intrinsic BHZ hamiltonian $\mathcal{H}_0$ 
captures the essential topological features of the QSH phase:  
\begin{eqnarray}
\label{h0}
\mathcal{H}_0 (k_x,k_y)= 
m_{\textbf{k}} s_0\tau_z + v_x k_x s_z\tau_x -  v_y k_y s_0 \tau_y, 
\end{eqnarray}
where $s_{x,y,z}$ and $\tau_{x,y,z}$ are Pauli matrices for the spin and orbital degrees of freedom respectively, and $s_0 = \mathbb{I}_{2\times 2}$. Here $m_{\textbf{k}} = M - C_x k_x^2 - C_yk_y^2$ with $M,C_x,C_y>0$, $2M=1\,{\rm eV}$ captures the strong topological band inversion found in monolayer WTe$_2$, and $v_{x,y}/\hbar$ are the Dirac velocities along $x$ and $y$ directions, see Fig.~\ref{fig1}a. 

$\mathcal{H}_{R,I}$ arise when inversion symmetry is broken and do not alter the global topological features of $\mathcal{H}$. Rashba $\mathcal{H}_R$ couple spin blocks and Ising $\mathcal{H}_I$ mix the orbital textures~\cite{likun,berry_pablo}:
\be
\mathcal{H}_R = -\delta_x s_x \tau_y, \quad \mathcal{H}_I = -\delta_z s_z \tau_y, 
\label{hir}
\ee
where $\delta_{z,x}$ describe the strength of Ising and Rashba spin-orbit coupling respectively. 

We note that inversion breaking (manifest in $\mathcal{H}_{R,I}$) in monolayer WTe$_2$ can originate from a variety of sources that include for e.g., an applied electric field \cite{berry_pablo,likun}, coupling with the substrate, edge electric fields, or even a buckling of monolayer WTe$_2$ into a T$_d$ phase \cite{berry_pablo}; recently, inversion breaking in monolayer WTe$_2$ has been detected via photocurrent imaging~\cite{berry_pablo}. Regardless of its origin, these mix the spin sectors on Eq.~(\ref{h0}) and critically impact edge magnetoresponse \cite{jps_review, tarasenko,magnetoconductance, goldman_smith, zhang_iop}.
 
In order to construct the edge Hamiltonian from the topological band inversion encoded in Eq.~(\ref{h0}),
we examine an edge along the $x$-direction (Fig.~\ref{fig1}a), where monolayer WTe$_2$ electrons occupy $y\geq0$; $y<0$ is the vacuum. For each $k_x$, two topological edge states emerge with a gapless spectrum traversing the bulk bandgap when $\vec B=0$~\cite{kane_mele, bhz, jps_review, shen_book, finite_size, tarasenko}; these can be directly obtained by an exact numerical solution (ENS) of the coupled partial differential equation in $\mathcal{H}$ when $k_y \to - i\partial_y$, see {\bf SI}. 

To clearly exhibit the role $\mathcal{H}_{R,I}$ plays, however, we analyze the structure of the edge wavefunctions. In so doing, we write the 
edge zero modes, $|\Psi_s \ra$: 
\be
\label{eq:ansatz}
\la \vec r |\Psi_{s}\ra = \sum_{n} a_{{n}}^s {\rm exp}({-y/\lambda_{{n}}^s})|u_n^s\ra, 
\ee
where $s=\pm 1$, $\lambda_{n}^s$ are decay lengths of the edge state into the bulk, and $|u_n^s\ra$ are pseudospins capturing the relative spin and orbital composition of the zero modes; these are obtained from solving $[\mathcal{H}(0,i\lambda_{n}^s)] |u_n^s\ra=0$. 
Here the index $n=\pm 1$ arises from the quadratic $m_{\vec k}$ dependence, and $a_{+1}^s = -a_{-1}^s $ are normalization constants that ensure the wavefunction vanishes at $y=0$ as well as far form the edge $\Psi(y=0)=\Psi(y\rightarrow\infty) = 0$.

Using the edge zero modes in Eq.~(\ref{eq:ansatz}) we directly construct the edge hamiltonian 
by projecting the bulk hamiltonian $\mathcal{H}_0(k_x,0)$ onto the zero modes along the edge [Eq.~(\ref{eq:ansatz})]. 
We find the
eigenstates of the edge Hamiltonian, $\{ |\Phi_1 \ra, |\Phi_2 \ra \}$, are: 
\be
\label{eq:basis}
\begin{pmatrix}
  |\Phi_1 \ra\\
  |\Phi_2 \ra
\end{pmatrix} = \begin{pmatrix}
   \cos\frac{\chi_1}{2} e^{-i \chi_2/2} & \sin\frac{\chi_1}{2} e^{i \chi_2/2}  \\
   \sin\frac{\chi_1}{2} e^{-i \chi_2/2}  & -\cos\frac{\chi_1}{2} e^{i \chi_2/2} 
\end{pmatrix}
\begin{pmatrix}
   | \Psi_{+} \ra\\
   | \Psi_{-} \ra
\end{pmatrix}. 
\ee
In this basis, $\{ |\Phi_1 \ra, |\Phi_2 \ra \}$, the edge Hamiltonian is diagonal and can be written as the first term of Eq.~(\ref{eq:edge}). Keeping only leading order terms in $1/M$, we have ${\rm tan} \chi_1 \approx {\rm sgn}(\delta_x)|\Gamma|/\gamma$ and ${\rm tan}\chi_2 \approx {\rm Im}[\Gamma]/{\rm Re}[\Gamma]$ where ${\rm Re}[\Gamma] \approx \delta_x [1/(\delta) - \delta C_y/(v_y^2 M)] $ and ${\rm Im}[\Gamma] \approx \delta_x[(v_y^2 M - \delta^2C_y)/(4M^3 C_y)]$ is controlled by Rashba SOC, and $\gamma \approx \delta_z/\delta$  is controlled by Ising coupling SOC. The zero modes $| \Psi_{\pm} \ra$ on the RHS of Eq.~(\ref{eq:basis}) possess pseudo-spinors that read (to leading order) as 
\be
|u_n^s\ra = \mathcal{N}_s\left(\Sigma_{s}, \Sigma_{s},1 ,1\right)^T,
\quad \Sigma_{s} = (\delta_z + s\delta)/\delta_x, 
\label{eq:zeromode}
\ee
where $\mathcal{N}_{+} =[2(\Sigma_s^2 +1)]^{-1/2} $ and  $\mathcal{N}_{-} ={\rm sgn}(\delta_x) [2(\Sigma_s^2 +1)]^{-1/2} $ and $\delta=\sqrt{\delta_x^2 + \delta_z^2}$ with
\begin{equation}
[\lambda_{n}^s]^{-1} =  \frac{v_y}{2C_y} +\frac{n\delta}{2\sqrt{MC_y}} - i n s \sqrt{\frac{M}{C_y}} +\sqrt{\frac{M}{C_y}}\mathcal{O}\left(\eta^2\right), 
\label{eq:decay} 
 \end{equation}
where dimensionless $\eta=v_y/(2\sqrt{MC})$ is small due to the strong topological band inversion in monolayer WTe$_2$. Using typical parameters for monolayer WTe$_2$ we find $\eta = 0.15$ is small. At $\vec B=0$, $\{ |\Phi_1 \ra, |\Phi_2 \ra \}$ in Eq.~(\ref{eq:basis}) compose the gapless QSH edgestates that propagate with renormalized edge velocity $\tilde{v} = v_x\sqrt{\gamma^2 + |\Gamma|^2}$, see Eq.~(\ref{eq:edge}). 

{\it Cooperative spin-canting -- } The spin orientation of the edge states is directly controlled by $\delta_x, \delta_z$. For example, when $\delta_z, \delta_x = 0$, $\{ |\Phi_1 \ra, |\Phi_2 \ra \}$ in Eq.~(\ref{eq:basis}) are eigenstates of $s_z$: $\{(1,1,0,0)^{T},(0,0,1,1)^{T}\}$. When $\delta_z, \delta_x \neq 0$, however, $\{ |\Phi_1 \ra, |\Phi_2 \ra \}$ can in general cant away from the poles of a Bloch sphere (where north/south correspond to spin up/down). Such rotation of edge spin orientations are readily found in the familiar HgTe QSH systems, where small $\delta_x$ leads to significant canting of spins on the edge since $M \sim $ several meV is small in those systems~\cite{tarasenko}.~In contrast, large $2M= 1\, {\rm eV}$ in WTe$_2$ suppresses the power of $\mathcal{H}_R$ or $\mathcal{H}_I$ {\it individually} in canting the edge spin orientation. Indeed, when $\delta_z = 0$ (so that $\chi_1 = 90^\circ$) and $\delta_x \neq 0$ the spin orientations barely cant away from the north/south pole since $\chi_2 \approx 2 (\delta_x/M) \eta^2 \sim 0.005 \, {\rm rad} = 0.3^\circ$ using $\delta_x \approx 50 \,{\rm meV}$. Similarly, for $\delta_z \neq 0$ but $\delta_x =0$, the edge spins continue to be aligned along $s_z$.  

Instead, when {\it both} $\delta_x,\delta_z \neq 0$ a cooperative effect ensues to produce a large spin canting which is relatively insensitive to $M$. For typical values of $\delta_x,\delta_z$ in WTe$_2$ we find $|{\rm tan} \chi_1| \sim$ unity signaling significant rotation away from $\vec s_z$. As we now discuss, this departure (in how the WTe$_2$ edge spin orientation behaves) from the more familiar case of HgTe/CdTe leads to COM and a distinctly different edge magnetoresponse.

\begin{figure}[t!]
\includegraphics[width=\columnwidth]{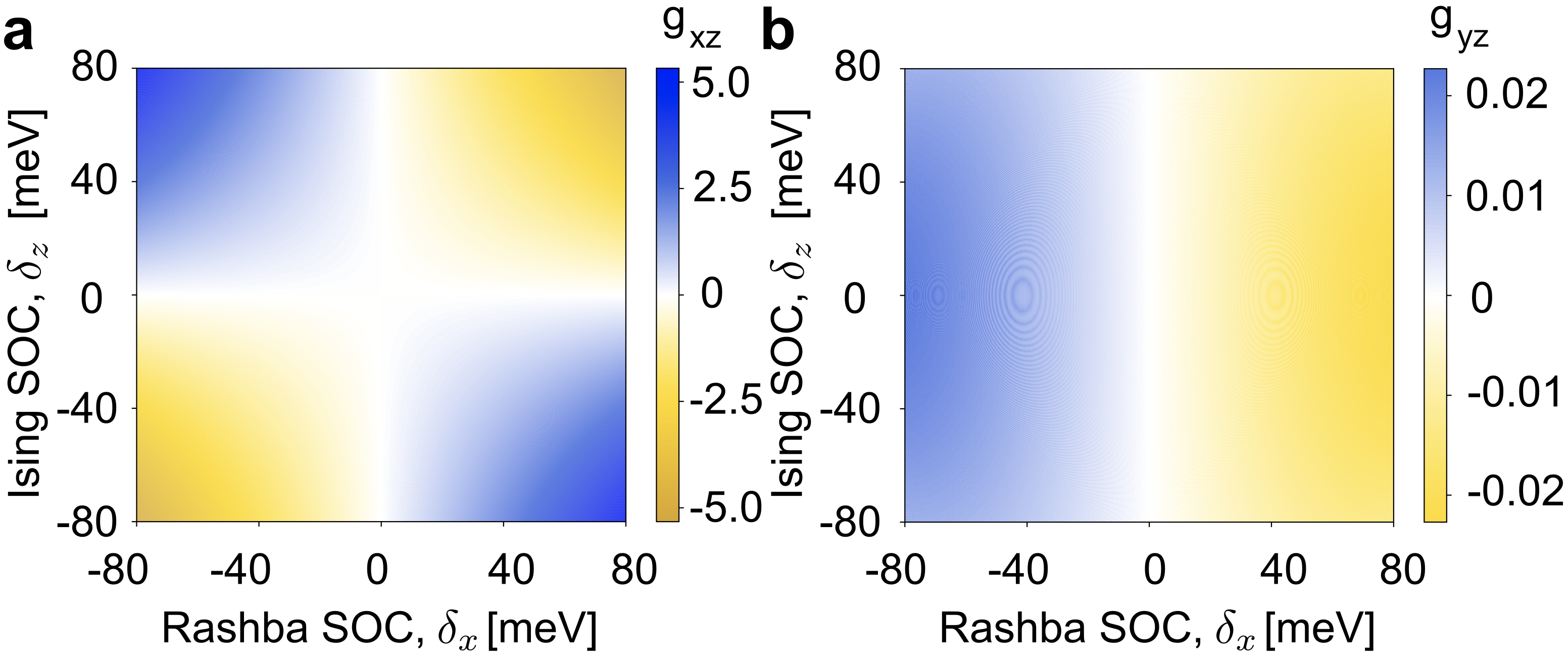}
\caption{Effective gyromagnetic coefficients for a perpendicular magnetic field for the edge parallel to mirror axis ($x$-axis) obtained from numerical evaluation of edge eigenstates in Eq.~(\ref{eq:basis}), see text. Panel (a) shows gyromagnetic coefficient corresponding to COM, $g_{xz}$ which arises from the combined action of finite Rashba and Ising SOC, and panel (b) displays $g_{yz}$, the gyromagnetic coefficient that corresponds to an ordinary orbital response~\cite{jps_review, magnetoconductance, tarasenko}. The latter only requires a non-zero Rashba SOC. Strikingly, $g_{xz}$ (panel a) is two orders of magnitude larger than $g_{yz}$ (panel b). Here we have used the same WTe$_2$ parameters as Fig.~1}.
\label{f2}
\end{figure}

{\it Cooperative edge magnetic moment -- } We first concentrate on the edge orbital magnetoresponse which is particularly sensitive to 
spin orientation. Orbital motion can be described via minimal coupling in the bulk as $\mathcal{H}_0(-eyB_z/\hbar c,0)$ where we have used a Landau gauge. Using $\{ |\Phi_1 \ra, |\Phi_2 \ra \}$ basis, we analyze the effect of the orbital motion on the edge electrons via $\bra{\Phi_1} \mathcal{H}_0(-eyB_z/\hbar c,0) \ket{\Phi_2}$. This produces $B_z$ induced terms in Eq.~(\ref{eq:edge}) that gap the edge spectrum, namely $g_{xz}$ and $g_{yz}$; these terms do not commute with $\sigma_z$. $g_{xz}$ and $g_{yz}$ are shown in Fig. 2a,b obtained by computing the above matrix element with a numerical solution of Eq.~(\ref{eq:basis}) keeping all orders.

Strikingly, Fig. 2a,b reveals that $g_{xz}$ is more than a $100$ times larger than $g_{yz}$. Further, while $g_{yz}$ is finite so long as $\delta_x \neq 0$, $g_{xz}$ arises only when {\it both} $\delta_x,\delta_z \neq 0$. The dichotomy in magnitudes and behavior of $g_{xz}$ and $g_{yz}$ vividly display a {\it cooperative} effect: in the presence of both Ising and Rashba SOC, large orbital gaps can be opened by $B_z$ [from $g_{xz}$]; in contrast, orbital response is severely suppressed when $\delta_x \neq 0,\delta_z= 0$ [from $g_{yz}$].

We identify the $\delta_x,\delta_z \neq 0$ cooperative behavior of $g_{xz}$ as COM. Importantly, sizeable COM persists even as $M$ overwhelms $\delta_x,\delta_z$. This large COM value directly proceeds from the strong canting of spins in the edge eigenstates when both $\delta_x,\delta_z \neq 0$, as discussed above. To see this link explicitly, we analyze the $M$ dependence of $g_{xz}$ directly, keeping only the leading order terms in Eq.~(\ref{eq:basis}): 
\be
g_{xz} \approx -\frac{A\delta_x\delta_z}{M}  = -1.62 \frac{(\delta_x [{\rm meV]/50) (\delta_z [{\rm meV}]/50)}}{(M \, [{\rm meV}]/500)}, 
\label{eq:gxz}
\ee
where $A=8m_e v_x C_y^2/(\hbar^2 v_y^3)$ with $m_e$ the free electron mass, and in the second line we have used typical ranges of fitted values for WTe$_2$ band parameters (see {\bf SI}). Comparing $g_{xz}$ in Fig.~\ref{f2}a and Eq.~(\ref{eq:edge}), we find the strong canting of spins enables COM magnitude of several Bohr magnetons for WTe$_2$. In contrast, $g_{yz} \sim 0.02$ (Fig.~\ref{f2}b) is highly suppressed, and scales as $\delta_x/M^2$ mirroring the small $\chi_2$ canting in the edge spins (suppressed by large $M$) in much the same fashion as that found for the edge orbital moments of HgTe~\cite{jps_review, tarasenko}.

While the cooperative spin canting mechanism and COM are general effects, we expect COM to be especially pronounced in monolayer WTe$_2$ due to its misaligned Te atoms in the top and bottom layers (see Fig.~\ref{fig1}a). This misalignment inextricably link Rashba (arising from out-of-plane dipole) and Ising (from in-plane dipole) SOC~\cite{berry_pablo,likun}. Further, the large $M$ of monolayer WTe$_2$ severely suppresses the ordinary orbital response that arises from $H_R$ alone (i.e. independent of $\delta_z$) that is typically found in small $M$ systems such as HgTe quantum wells~\cite{jps_review, magnetoconductance,tarasenko}. Indeed, COM (Fig.~\ref{f2}a) is consistent with the sizeable out-of-plane edge gap opening recently measured in WTe$_2$ monolayers~\cite{qsh_wte2_pablo,qsh_wte2_david}. 

To obtain the full edge magnetoresponse, we now also include the (pure spin) Zeeman effect where magnetic field directly couples with $\vec s$ in the bulk. These only contribute to the diagonal terms of Eq.~(\ref{eq:edge}) and do not contribute to gap opening along the edge when $\vec B = B_z \hat{\vec{z}}$~\cite{magnetoconductance, tarasenko, ronny_zeeman}, see {\bf SI} for full discussion.
Taking typical values of bulk electronic Lande g-factors $\approx 2$ and projecting onto the edge, we obtain $g_{xx}, g_{yy} \sim 1.2 - 2.0$ which weakly depend on $\delta_x,\delta_z$, {\bf SI}. Combining orbital (off-diagonal) and spin Zeeman (diagonal) terms in Eq.~(\ref{eq:edge}), we obtain an anisotropic B-field induced edge (full) gap, $\Delta (\vec B)$, Fig.~\ref{fig1}b,c. 

Crucially, for most orientations of $\vec B$ (when azimuthal $\varphi \neq 90^\circ$), we find that $\Delta (\vec B)$ is minimized at an oblique polar angle: i.e. $\theta$ is neither zero or $90^\circ$. This is in stark contrast to that found in HgTe QSH systems where minimal gap occurs when $\theta= 90^\circ$~\cite{jps_review}. 
Strikingly, when $\vec B$ lies in the $x$-$z$ plane (azimuthal $\varphi=0$), $\Delta (\vec B)$ nearly vanishes at a critical angle $\theta_c$ (see blue curve in Fig.~1b); indeed the residual gap $\theta_c$ is $3.5 \, \mu {\rm eV}$ (obtained from Fig.~\ref{f2}) -- two orders of magnitude smaller than the maximum gap opening (Fig.~1b). Noting that $g_{xz} \gg g_{yz}$ and specializing to $\vec B = |\vec B| (\sin \theta,0,\cos\theta)$ in the $x$-$z$ plane, we find 
\begin{equation}
\label{eq:gap}
\Delta (\vec B) \approx g_{\rm eff}  \mu_B |\vec B| 
\big|{\rm sin}(\theta - \theta_c)\big|, \quad {\rm tan} \theta_c = - \frac{g_{xz}}{g_{xx}},
\end{equation}
where $g_{\rm eff} = \sqrt{g_{xx}^2 + g_{xz}^2}$. 
Taking $\delta_z = 70$ meV and $\delta_x = 40$ meV as a illustration, we obtain $\theta_c = 51^\circ$ (Fig.~1b). 
$\Delta (\vec B)$ exhibits a $180^\circ$ periodicity; the next zero in Eq.~(\ref{eq:gap}) occurs at $\theta= \theta_c + 180^\circ$. Maximal $\Delta (\vec B)$ occurs when $\theta = \theta_{\rm max} = \theta_c + 90^\circ$. We note that when $\vec B$ is directed in planes other than $x$-$z$ (i.e. $\varphi \neq 0$) such near vanishing of gap does not occur (see e.g, Fig.~1c, green and red curves), and is non-vanishing and sizeable in all directions. 

While we have concentrated on edges along the $x$ direction, terminations along other edges can influence the edge behavior~\cite{anton}. 
For generic edges, we find COM and 
angle-dependent $\Delta(\vec B)$ persists, with a well-defined $\widehat{\vec{B}}_*$ along which magnetic field is ineffective at gapping the edge spectrum, see {\bf SI}. We note that while the precise value of $\theta_c$ along generic edge terminations (i.e. not $x$-edge) can vary and depart from that described in Eq.~(\ref{eq:gap}), COM and its effect on magnetoresponse behavior (namely $\widehat{\vec{B}}_*$) nevertheless proceeds directly from the cooperative $\delta_x$, $\delta_z$ induced spin canting along the edge.

\begin{figure}
\includegraphics[width=\columnwidth]{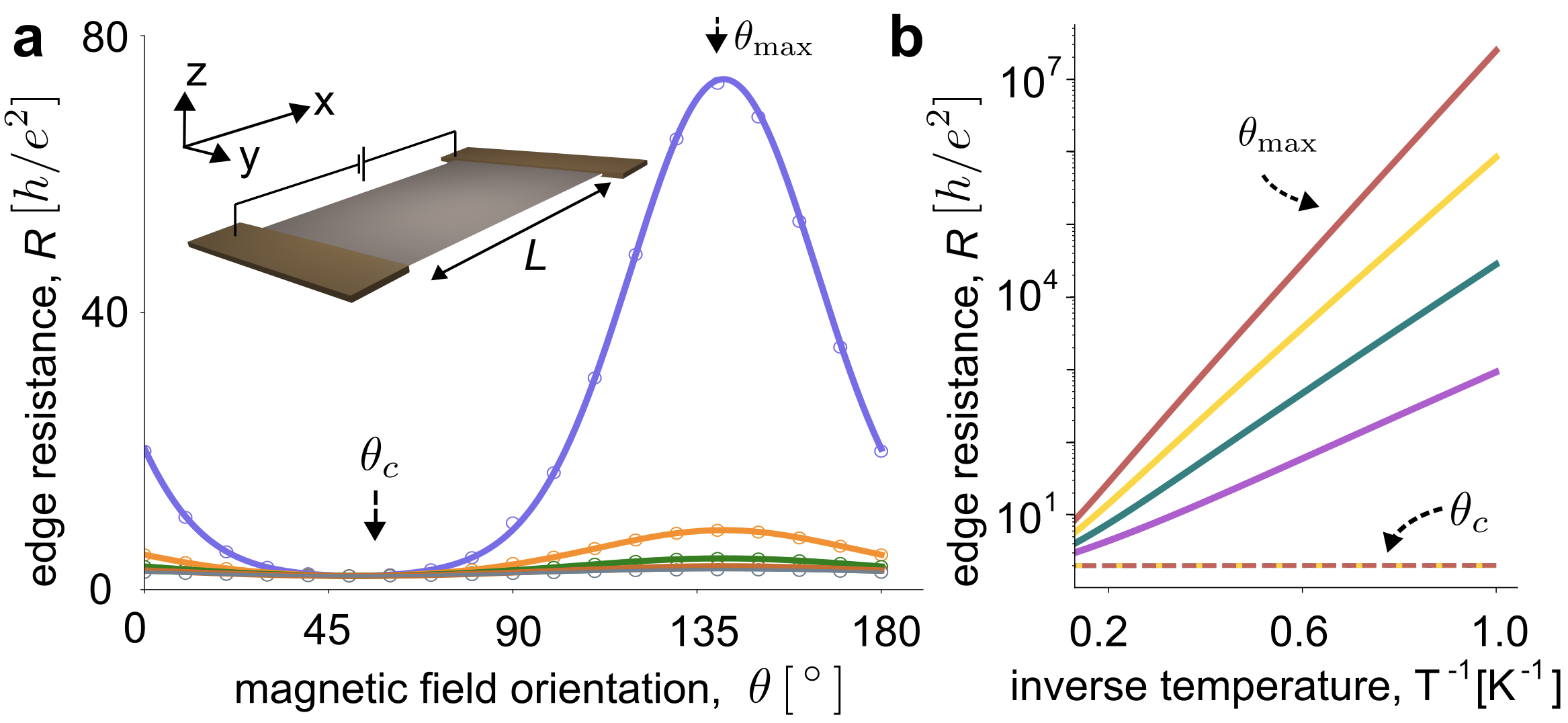}
\caption{Edge magnetoresistance $R (\vec B)$ for magnetic field directed in $x$-$z$ plane, i.e. $\textbf{B} = |\textbf{B}|(\sin\theta, 0, \cos\theta)$. (a) $R(\vec B)$ varies with $\theta$ at different temperatures $T = 2, 4, 6, 8, 10 \, {\rm K}$ (top to bottom) and at a fixed $|\textbf{B}|  = 7\, {\rm T}$. $R(\vec B)$ displays no temperature variation at $\theta_c$; in contrast, $R(\vec B)$ exhibits large temperature changes at $\theta_{\rm max}$. Solid lines are obtained using edge Hamiltonian in Eq.~(\ref{eq:edge}), circles are obtained from edge spectrum ENS (see {\bf SI}) displaying good agreement. (b) $R(\vec B)$ taken at $\theta_{\rm max}$ (solid curves) display thermally activated behavior 
with slope (in logarithmic scale) increasing as $|\vec B| = 6, 8, 10, 12 \, {\rm T}$ increases (bottom to top). In contrast, $R(\vec B)$ at $\theta_c$ (dashed lines) does not change with temperature or with magnetic field. In all plots we have used device parameters $L = 500 \, {\rm nm}$, $\ell_0 = 100\, {\rm nm}$, $R_c = h/2e^2$ and a chemical potential fixed at $\mu = 0.3 \, {\rm meV}$ as an illustration. The other parameters used are the same as in Fig.~\ref{fig1}.}
\label{fig3}
\end{figure}

{\it Edge magnetoresistance -- }  
The gap opening in the edge state spectrum $\Delta (\vec B)$ when magnetic field is applied 
directly impacts edge transport producing an edge magnetoresistance. We track the resistance along the edge via $R (\vec B) = R_c + R_s (\vec B)$, where $R_c$ is the contact resistance while $R_s$ is the resistance of edge channel. The latter can be computed via $R_s (\vec B)= L \rho (\vec B) $, where $L$ is the length of the channel and $\rho (\vec B)$ is the resistivity of the edge channel \cite{ashcroft, lundstorm}: $[\rho(\vec B)]^{-1} = e^2\sum_{k_x} (\partial \varepsilon/\hbar\partial k_x)^2 \tau(\varepsilon)(-\partial_\varepsilon f)$ where $\varepsilon (k_x)$ are the eignenergies of Eq.~(\ref{eq:edge}), 
$f = \{1+ {\rm exp} [\beta (\varepsilon- \mu)]\}^{-1}$ is the Fermi function and $[\tau(\varepsilon)]^{-1} = (2\pi/\hbar) \nu(\varepsilon) |\bra{\psi(-k_x)}V(x)\ket{\psi(k_x)}|^2$ is the relaxation rate. Here $\psi(k_x)$ are the eigenstates of Eq.~(\ref{eq:edge}), $\nu(\varepsilon)$ the density of states of the 1D edge channel, $\mu$ the chemical potential, $\beta = 1/k_BT$, and $V(x)$ is the impurity potential. 
In what follows, we choose short range impurities $V(x) = \sum_j u \delta(x-x_j)$ as an illustration with $u$ the strength of a single impurity and $j$ sums over all impurities in the channel. The disorder average $\la V(x) V(x')\ra = n_{{\rm imp}} u ^2\delta(x-x')$ where $n_{\rm imp}$ is the impurity density yielding a length scale $\ell_0 = \tilde{v}^2/(\mathcal{U}^2 n_{{\rm imp}})$ \cite{ajit, giovannie}. \footnote{The mean free path of a 1D electron system with left and right movers can be written as $\ell = \ell_0 / | \la\psi(-k_x) |\psi (k_x)\ra |^2$. For a helical 1D liquid at $\vec B=0$, $\ell$ diverges since the left- and right-movers are time-reversed pairs yielding $\la \psi (-k_x) |\psi (k_x)\ra$ that vanishes. However, when $\vec B \neq 0$, this protection is lifted, and $\la \psi (-k_x) |\psi (k_x)\ra \neq 0$ allowing finite mean free-paths. For completely non-helical liquids $\la \psi(-k_x) |\psi (k_x)\ra = 1$ yielding $\ell = \ell_0$ as expected for the ``non-helical'' mean free path.}  

In Fig.~\ref{fig3}a we plot the edge magnetoresistance $R(\vec B)$ for an edge along the $x$-direction for a fixed magnitude of magnetic field $|\vec B| = 7\, {\rm T}$ but applied at various orientations of $\vec B$ along the $x$-$z$ plane (i.e. $\varphi=0$), namely $\vec B = |\vec B| ({\rm sin}\theta, 0, {\rm cos}\theta)$. Edge magnetoresistance is large and highly sensitive to temperature close to $\theta =\theta_{\rm max}$ where the edge gap is largest; indeed, as temperature is lowered, the magnetoresistance climbs rapidly in an exponential fashion, Fig.~\ref{fig3}b. 

In contrast, close to $\theta = \theta_{\rm c}$ the edge magnetoresistance remains nearly constant with temperature since gap opening close to $\theta_{\rm c}$ almost vanishes, see blue curves in Fig.~\ref{fig1}b,c. Indeed, curves at various temperatures in Fig.~\ref{fig3}a,b collapse on each other at $\theta = \theta_{\rm c}$. The insensitivity of magnetoresistance to temperature provides a direct experimental means to determine $\widehat{\vec{B}}_*$ -- the direction wherein applied magnetic field is ineffective at gapping out the edge states. Away from the $x$-$z$ plane, i.e. $\varphi\neq0$, we note that resistance curves at various temperatures no longer collapse on each other for any value of $\theta$ (see {\bf SI}) since $\Delta(\vec B)$ is generically finite and sizable in all directions when $\varphi\neq0$. 

COM is a direct consequence of the cooperation between $\mathcal{H}_R$ and $\mathcal{H}_I$ that plays a critical role in both the edge spin structure and its magnetoresponse. Given the tight spatial confinement of the edge wavefunctions $\sim 2C_y/v_y = 1.9\, {\rm nm}$ for typical WTe$_2$ parameters, we expect that edge magnetoresponse $\mathcal{H}_{R/I}$ in the region close to the edge, where for e.g., strong confining edge electric fields will inevitably appear close to sample boundaries~\cite{glazman}. Additionally, while here we have concentrated on COM and its relation to $\mathcal{H}_R$ and $\mathcal{H}_I$, we remark that the cooperative spin canting mechanism -- arising from combined action of two types of SOC -- is general; we anticipate that such cooperative spin canting can also ensue with other SOC mechanisms.

Perhaps most striking, from a technological perspective, is COM magnetoresponse that exhibits an unusual near-zero of magnetoresistance when magnetic field is applied along the direction $\widehat{\vec{B}}_*$. This near-zero in magnetoresistance enables very large anisotropic magnetoresistance: comparing resistance at $10 \, {\rm T}$ for $\theta_{\rm max}$ and $\theta_c$ (Fig.~\ref{fig3}b) we expect giant anisotropic magnetoresistance of several million $\%$ at low temperature can be achieved in WTe$_2$. These are competitive with other large magnetoresistance materials at similar $\vec B$ and temperature ranges, (e.g., bulk 3D topological semimetals~\cite{ong}, or 3D bulk charge compensated transition metal dichalcogenides \cite{ali}).

\vspace{2mm} 
{\bf Acknowledgments} - We thank David Cobden, Wenjin Zhao, and Valla Fatemi for useful conversations. This work was supported by the Singapore National Research Foundation (NRF) under NRF fellowship award NRF-NRFF2016-05, a Nanyang Technological University start-up grant (NTU-SUG), and Singapore MOE Academic Research Fund Tier 3 Grant MOE2018-T3-1-002.

\clearpage
\newpage

\renewcommand{\theequation}{S\arabic{equation}}
\renewcommand{\thefigure}{S\arabic{figure}}
\renewcommand{\thetable}{S\Roman{table}}
\makeatletter
\renewcommand\@biblabel[1]{S#1.}
\setcounter{equation}{0}
\setcounter{figure}{0}

\appendix
\onecolumngrid
\section{Supplementary Information for ``Cooperative orbital moment and edge magnetoresistance in monolayer WTe$_2$"}
\twocolumngrid

\section{Symmetry and WTe$_2$ low-energy hamiltonian}
The low-energy electronic excitations for monolayer WTe$_2$ with inversion symmetry can be captured by a minimal four-band BHZ Hamiltonian
\begin{eqnarray}
\label{hs0}
\mathcal{H}_0 (k_x,k_y)= \epsilon_{\textbf{k}}\mathbb{I} + m_{\textbf{k}} s_0\boldsymbol{\tau}_z + v_x k_x s_z\tau_x - v_y k_y s_0 \tau_y, 
\end{eqnarray}
where $\epsilon_{\textbf{k}} = (\epsilon_c + \epsilon_v)/2$ is the dispersive part of the Hamiltonian and $m_{\textbf{k}} = (\epsilon_c - \epsilon_v)/2$ is the momentum dependent mass capturing the topological inversion. Here, $\epsilon_i = c_{i,0} - c_{i,x}k_x^2 - c_{i,y}k_y^2$, $i\in\{c,v\}$ is the dispersion for conduction and valence band respectively. Taking 
$c_{c,0}=1$ eV and $c_{v,0}=0$ eV  gives the magnitude of topological band inversion in Eq.~(\ref{hs0}):
$2M = 1$ eV~\cite{likun}. Throughout the text we use values for $v_x = 1.71$ eV\AA  ~and $v_y = 0.48$ eV\AA~ obtained from a model fitting with available ARPES data~\cite{berry_pablo,qsh_tmds}. We have chosen $(C_x, C_y) = (5.5,4.5)$ eV\AA$^2$ parameter values used in the main text that are obtained by L\"owdin partitioning~\cite{likun} and lie in the typical range for $C_x, C_y$. Lastly, we note that since $ \epsilon_{\textbf{k}} \mathbb{I} $ is a scalar, contributing to spin/orbital blocks equally it does not affect the structure of the edge state wavefunctions. For brevity as well as clarity, below and in the main text, we have suppressed $\epsilon_{\textbf{k}}$ dependence.

When inversion symmetry is broken in monolayer WTe$_2$, Ising and Rashba SOC arise~\cite{likun} and can be captured via
\be
\mathcal{H}_R = -\delta_x s_x \tau_y, \quad \mathcal{H}_I = -\delta_z s_z \tau_y, 
\label{hsir}
\ee
where $\delta_{x(z)}$ is the Rashba (Ising) SOC strength. Here $\mathcal{H}_R $ couple the spin blocks, and $\mathcal{H}_I$ mix its spin texture. In the main text, and unless stated otherwise, we use parameter values for $\delta_z = 70$ meV and $\delta_x = 40$ meV. These are within the range of SOC strengths achievable in WTe$_2$~\cite{berry_pablo}. Together with $H_0$ above, the full hamiltonian at zero magnetic field is $\mathcal{H} = \mathcal{H}_0 + \mathcal{H}_R + \mathcal{H}_I$. 

\subsubsection{Emergent symmetries}

We now turn to discussing the symmetries present in our system. First we note, that without external field, pristine WTe$_2$ monolayer crystal possess time-reversal (TR) symmetry and a (crystallographic) point group symmetry $P2_{1}/m$ that contains four symmetry operations, which include inversion symmetry ${\cal I}: \vec r \to - \vec r$ and mirror symmetry ${\cal M}_y : y \to - y$~\cite{likun}. These can be used to constrain and construct the low-energy hamiltonian that describes the long-wavelength electronic excitations~\cite{likun}. 

Crucially, we note that the low energy model $\mathcal{H}_0 (k_x,k_y)$ in Eq.~(\ref{hs0}) contains further {\it emergent} symmetries ${\cal M}_x : x \to - x$ and ${\cal M}_z : z \to - z$, under which the Hamiltonian $\mathcal{H}_0 (k_x,k_y) $ remains invariant. These emergent symmetries arise in the long wavelength description of monolayer WTe$_2$. Since they are spatial symmetries, they are valid even when TRS is broken.

Interestingly, the mirror operations ${\cal M}_x : x \to - x$ and ${\cal M}_z : z \to - z$ can be used to characterize the Ising and Rashba SOC in Eq.~(\ref{hsir}); both terms generically arise when inversion is broken. Specifically, $\mathcal{H}_{R(I)}$ is odd under ${\cal M}_{z(x)}$. This means that $\mathcal{H}_{R(I)}$ vanishes when the system is symmetric under ${\cal M}_{z(x)}$. Therefore nonvanishing $\mathcal{H}_{R(I)}$ directly proceed broken ${\cal M}_{z(x)}$ (when either [or both] is broken, inversion symmetry is broken), and can be thought to arise from out-of-plane (in-plane) dipole moment. This is consistent with their microscopic origin as matrix elements of the relativistic spin-orbit interaction: ${\hat H}_{\rm so} (\vec k) \sim ( \vec k + {\hat {\vec p}}/m_0 )\cdot \vec s \times \nabla \phi (\vec r)$~\cite{winkler_book}.

A similar analysis can also be applied for the QSH edge states in WTe$_2$. In the presence of an applied magnetic field, the QSH edge states along the $x$-edge can be generically described via Eq.~(\ref{eq:edge}) of the main text. Here we reproduce it for the convenience of the reader: 
\be
\label{eq:edge-SI}
H^{\rm edge} (k_x) = \tilde v \sigma_z k_x + \frac{\mu_B}{2}\sum_{ij} g_{ij} \sigma_{i} B_j, 
\ee
where we have taken the lowest symmetry allowed terms, neglecting terms higher order in $k$ and $B$. Note that the first term in Eq.~(\ref{eq:edge-SI}) is symmetric under both ${\cal M}_{z}$ and ${\cal M}_{x}$ operations.

We now concentrate on the second term of Eq.~(\ref{eq:edge-SI}). Under the mirror operation ${\cal M}_{z}$, we have 
\begin{align}
\sigma_{x,y} \to - \sigma_{x,y}, 
&\quad
\sigma_{z} \to \sigma_{z},
\nn
B_{x,y} \to - B_{x,y} ,
&\quad
B_{z} \to B_{z} .
\end{align}
since $\boldsymbol{\sigma}$ and $\vec B$ are pseudo-vectors. This means that when the system is symmetric under ${\cal M}_{z}$, the out-of-plane $B$-field terms $g_{xz} \sigma_x B_z$ and $g_{yz} \sigma_y B_z$ must vanish.

We note that a further constraint can be gleaned by analyzing the behavior of Eq.~(\ref{eq:edge-SI}) under the mirror operation ${\cal M}_{x}$. In the same fashion as above, this operation flips $\vec B$ and $\boldsymbol{\sigma}$ perpendicular to $x$. This means that when the system is symmetric under ${\cal M}_{x}$, the $B$-field terms $g_{xz} \sigma_x B_z$, $g_{xy} \sigma_y B_x$, and $g_{yx} \sigma_x B_y$ must vanish. When both ${\cal M}_{z}$ and ${\cal M}_{x}$ are symmetries of the system, all off-diagonal terms in $g_{ij} \sigma_i B_j$ must vanish. Similarly, we note that while $g_{yz} \sigma_y B_z$ can manifest when ${\cal M}_{z}$ symmetry is broken but ${\cal M}_{x}$ symmetry is preserved. $g_{xz} \sigma_x B_z$ requires both ${\cal M}_{x}$ and ${\cal M}_{z}$ to be broken. This coincides with our analysis of COM in the main text that shows large $g_{xz}$ arising when both $\delta_x,\delta_z \neq 0$.

\subsection{Quantum spin Hall edge states in WTe$_2$}

The QSH edge states along an $x$-edge analyzed in the main text proceeds directly from the topological band inversion in Eq.~(\ref{hs0}). In so doing we consider the following geometry: monolayer WTe$_2$ electrons occupy $y\geq0$; $y<0$ is the vacuum. The edge state spectrum and wavefunction spinor and spatial profile can be directly computed from $\mathcal{H}$ by setting $k_y \to -i \partial_y$ and solving for self-consistent wavefunction solutions. There are two principal methods to obtain the QSH edge states: (i) through an exact numerical solution (ENS) of the edge state spectrum by solving the coupled (in spin/orbital space) Schr\"odinger equations and (ii) constructing the edge hamiltonian from the zero modes of $\mathcal{H}$. As shown in the main text, we perform both methods which display good agreement with each other. In this section, we describe both methods in more detail.

\subsubsection{Exact numerical solution of edge state spectrum}

We first describe the ENS method. The edge states at zero field can be obtained from an ENS of a set of coupled Schr\"odinger equations with $k_y \to -i \partial_y$ and appropriate boundary conditions. Including both the effect of orbital motion of electrons (through minimal coupling in the landau gauge $k_x \to k_x - eyB_z/\hbar c$) as well as a (spin) Zeeman effect found below in Eq.(\ref{eq:zeemanSI}) we write the coupled Schr\"odinger equations as
\begin{equation}
\label{eq:numerical}
\left[\mathcal{H}\left(k_x -\frac{eyB_z}{\hbar c} , -i\p_y\right) + \mathcal{H}_Z(\textbf{B})\right]\boldsymbol{\xi}_{k_x}(y) = \epsilon\boldsymbol{\xi}_{k_x}(y)
\end{equation}
for each $k_x$ along the $x$-edge. Here $\boldsymbol{\xi} (y) = \big[\xi_1(y), \xi_2(y), \xi_3(y), \xi_4(y)\big]^T$ is a general four component spinor wavefunction. Each of the $\xi_i(y)$ wavefunctions satisfy Dirichlet boundary conditions, $\xi_i(-L) = \xi_i (L) = 0$ on a strip geometry with $y\in[-L, L]$ and $L= 200$ nm. Here, as in the main text, $\vec B = |\vec B| (\sin \theta \cos \varphi, \sin \theta \cos \varphi, \cos \theta)$. 

\begin{figure}
\includegraphics[width = \columnwidth]{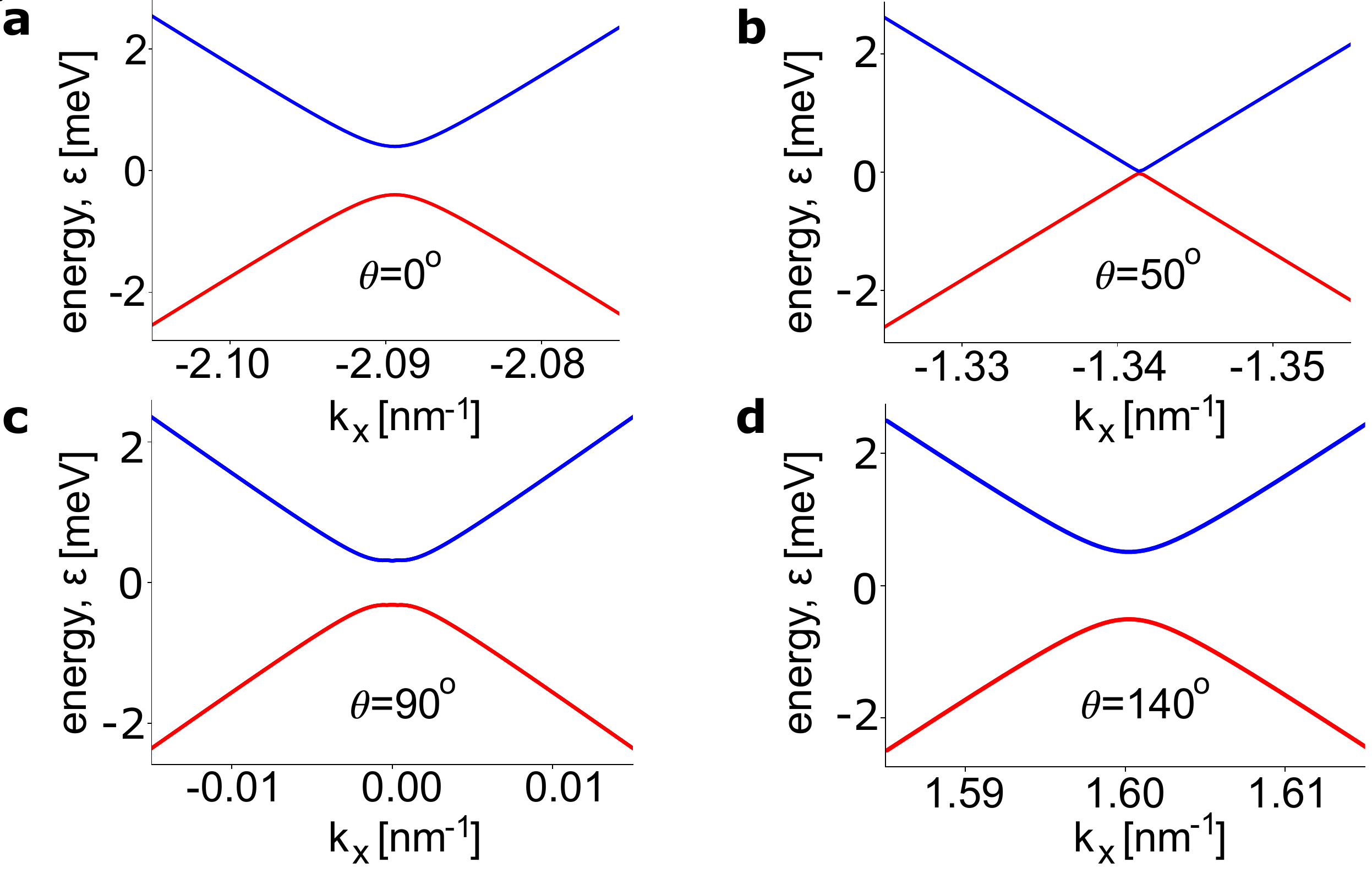}
\caption{(a,b,c,d) 
Edge spectrum generated from ENS algorithm (see text) by solving Eq. (\ref{eq:numerical}) for magnetic field $\textbf{B} = 7|(\sin\theta, 0, \cos\theta) $ T at $\theta = 0^{\circ}, 50^{\circ}, 90^{\circ}, 140^{\circ}$ respectively. Here, we use $\delta_x = 40$ meV and $\delta_x = 70$ meV for the strength of Rashba and Ising SOC respectively. } \label{spectrum}
\end{figure}

We solved Eq. (\ref{eq:numerical}) numerically using a standard numerical partial differential equation subroutine in {\it Mathematica} by utilizing finite element analysis.
The system of equations is solved along a line (along $y$-direction) of length $2L$ (composed of $2/10^{-3}$ elementary divisions) for every value of $k_x$ to obtain the edge spectrum. Sample edge spectrum as a function of $k_x$ along the $x$-edge are shown in Fig.~\ref{spectrum} for various values of $\theta$; here $\varphi=0$. 

In Fig.~\ref{fig1} and \ref{fig3} of the main text we display gap size and edge resistance obtained from the ENS edge state spectrum as open circles. In obtaining the gap size at each orientation of magnetic field $(\theta,\phi)$, we numerically obtained the minimum difference between upper (blue) and lower (red) bands in Fig.~\ref{spectrum} to obtain the full size of the gap. Note the near gap closure in the edge spectrum for $\theta = 50^\circ$ in Fig.~\ref{spectrum} in agreement with the critical angle $\theta_c $ found in the main text.

\subsubsection{Zero modes and constructing low-energy edge hamiltonian} 

In this section, we describe how to construct the low-energy edge hamiltonian from the zero modes of $\mathcal{H} (k_x, k_y \to -i\partial_y)$. Zero modes along the $x$-edge satisfy
\begin{equation}
\label{eq:oansatzs}
\la \vec r \ket{\Psi} = \sum_{j} a_{j} {\rm exp}(-y/\lambda_{j})\ket{u_j} {\rm exp}[ik_x x]
\end{equation}
with the boundary condition $\Psi(y=0)=\Psi(y\rightarrow \infty)=0$. Here $\lambda_j$ is the decay length associated with $j^{{\rm th}}$ decay mode, $|u_j\ra$ are 4 component spinors, and $a_j$ is the corresponding normalization coefficients. The zero-mode spinors and decay constants satisfy $\mathcal{H}(0,\lambda_j^{-1})\ket{u_j} = 0$. Solving for the zeros of ${\rm det} [\mathcal{H}(0,\lambda_j^{-1})]$, we obtain zero-mode decay lengths  
 \begin{equation}
 \label{eq:decay_org}
\lambda^{-1} = \pm \frac{v_y \pm \sqrt{v_y^2 - 4C_y (M\pm i \delta)}}{2C_y}, \quad \delta= \sqrt{\delta_x^2 + \delta_z^2}
 \end{equation}
that can take on eight possible values. Using the boundary condition $\Psi(y\rightarrow\infty)=0$, we discard four $\lambda^{-1}$ in Eq.~(\ref{eq:decay_org}) that have negative real parts (they are not normalizable in our space). The zero mode spinors that correspond to each $\lambda_j$ can then be obtained numerically; using these we construct two degenerate zero modes. 
The edge hamiltonian (and the numerical edge state wavefunction) can then be obtained in the standard fashion by projecting the bulk hamiltonian $\mathcal{H}(k_x,0)$ onto the space spanned by the degenerate zero modes in the same form as Eq.~(\ref{eq:edge}) of the main text. In our main text, these numerical edge state wavefunctions are employed in obtaining the orbital contributions to magnetoresponse (i.e. $g_{xz}$ and $g_{yz}$) shown in Fig.~2 of the main text. (Spin) Zeeman contributions, see below, are similarly obtained using these same edge state wavefunctions. For clarity we denote these numerical edge eigenstates (NE).

While the above procedure is standard and can be implemented numerically, to gain further insight into the structure of the edge wavefunctions, we now analyze approximate zero-modes that that can be expressed in a simple closed form. As shown in the main text, these allow us to expose the origin of COM -- cooperative spin canting. We do this by first noting that $\eta=v_y/(2\sqrt{MC_y}) = 0.15$ is small. Exploiting this small parameter, we expand $\lambda^{-1}$ in Eq.~(\ref{eq:decay_org}) in $\eta$ 
and obtain Eq. (\ref{eq:decay}) of the main text 
\begin{equation}
\left[\lambda_{n}^s\right]^{-1} =  \frac{v_y}{2C_y} +\frac{n\delta}{2\sqrt{MC_y}} - i n s \sqrt{\frac{M}{C_y}} +\sqrt{\frac{M}{C_y}}\mathcal{O}\left(\eta^2\right), 
\label{eq:decays} 
 \end{equation}
 where $\eta^2 = 0.02$ is small, and $s=\pm$ and $n=\pm$. 
In the same fashion as above, zero mode spinors can be obtained by solving the set of four simultaneous equations $\mathcal{H}(0,\left[\lambda_n^s\right]^{-1})\ket{u_s} = 0$, where $\left[\lambda_n^s\right]^{-1}$ is given by Eq. (\ref{eq:decays}). Using $M+C_y\left[\lambda_n^s\right]^{-2} = -2insM(\eta+n\Omega+\mathcal{O}(\eta^2,\Omega^2))$ and $v_y\left[\lambda_n^s\right]^{-1} = -2insM(\eta + \mathcal{O}(\eta^2, \eta\Omega))$, with $\Omega = \delta/(2M)$, the above system of equations yields the approximate zero-mode
spinors in Eq. (\ref{eq:zeromode}) of the main text:
\be
|u_s\ra = \mathcal{N}_s\left(\Sigma_{s}, \Sigma_{s},1 ,1\right)^T,  \quad \Sigma_{s} = (\delta_z + s\delta)/\delta_x, 
\label{eq:zeromodes}
\ee
where $\mathcal{N}_+ = [2(\Sigma_s^2 + 1)]^{-1/2}$ and $\mathcal{N}_- = {\rm sgn} (\delta_x) [2(\Sigma_s^2 + 1)]^{-1/2}$. We note, parenthetically, that the choice of zero modes is not unique. Indeed, other linear combinations of zero modes shown above are equally valid and produce the same physical observables and dependencies (e.g., edge gap behavior as a function magnetic field). 

Using Eq. (\ref{eq:decays}) and (\ref{eq:zeromodes}), we can write approximate edge zero mode wave functions (see also Eq.(\ref{eq:ansatz}) of the main text)
\be
\label{eq:ansatzs}
\la \vec r |\Psi_{s}\ra = \sum_{n} a_{{n}}^s {\rm exp}({-y / \lambda_{{n}}^s})|u_s\ra, 
\ee
where we have identified the $s= \pm 1$ with the spin degree of freedom for the two fold degenerate edge modes and $n$ represents the two-decay modes for each spin (note change in index $j$ to $n,s$ index).
Next, we determine the coefficients $a_n^s$ by using the other boundary condition $\Psi(y=0) = 0$, according to which $\sum a_n^s \ket{u_s} = 0$ is satisfied by $a_n^s = \{1,-1\}$. This leads to the degenerate edge wavefunctions 
$
\ket{\Psi_{s}} = \mathcal{K}\left[{\rm exp}\left(-y/ \lambda_+^{s}\right) - {\rm exp}\left(-y/ \lambda_-^{s}\right)\right]\ket{u_{s}}
$
with normalization constant $\mathcal{K} = \left[(v_y^2 M - \delta^2C_y)/(2v_y M C_y)\right]^{1/2}$.

The edge Hamiltonian can be obtained by projecting $\mathcal{H}_0(k_x,0)$ onto the zero modes in Eq.~(\ref{eq:ansatzs}) by computing the matrix elements 
$ \bra{\Psi_{\mu}}\mathcal{H}_0(k_x,0)\ket{\Psi_{\nu}}$, with $\mu,\nu \in s$ producing: 
\begin{equation}
\label{eq:edgeH}
H (k_x)= v_x k_x\left(\gamma\sigma_z + \textbf{Re}[\Gamma]\sigma_x + \textbf{Im}[\Gamma]\sigma_y\right),
\end{equation}
where ${\rm tan} \chi_1 = {\rm sgn}(\delta_x)|\Gamma|/\gamma$ and ${\rm tan}\chi_2 = {\rm Im}[\Gamma]/{\rm Re}[\Gamma]$ where ${\rm Re}[\Gamma] = \delta_x [1/(\delta) - \delta C_y/(v_y^2 M)] $ and ${\rm Im}[\Gamma] = \delta_x[(v_y^2 M - \delta^2C_y)/(4M^3 C_y)]$ is controlled by the Rashba coupling, and $\gamma = \delta_z/\delta$  is controlled by the Ising coupling. 

Eq. ~(\ref{eq:edgeH}) can can be diagonalized by the unitary operation $\mathcal{P}^{-1} H\mathcal{P} = \tilde{v}k_x \sigma_z$ producing the first term in Eq.~(\ref{eq:edge}) of the main text. Here we have used 
\begin{equation}
\label{eq:unitary_matrix}
\mathcal{P} = \begin{pmatrix}
   \cos\frac{\chi_1}{2} e^{-i \chi_2/2} & \sin\frac{\chi_1}{2} e^{-i \chi_2/2}  \\
   \sin\frac{\chi_1}{2} e^{i \chi_2/2}  & -\cos\frac{\chi_1}{2} e^{i \chi_2/2} 
\end{pmatrix}
\end{equation}
Further, the corresponding edgestate eigenfunctions can be similarly written as $\left(\ket{\Phi_1}, \ket{\Phi_2}\right)^T = \mathcal{P}^{-1} \left(\ket{\Psi_+},\ket{\Psi_-}\right)^T$, see also Eq.~(\ref{eq:basis}) of the main text.

\section{Magnetic field induced gap opening}

In this section, we discuss how magnetic field opens up a gap in the edge spectrum by analyzing how various both (i) the orbital motion of electrons, and (ii) a (spin) Zeeman interaction, can affect the structure of the edge eigenstates.

\subsubsection{Orbital magnetoresponse}

COM discussed in the main text can be directly obtained from analyzing the behavior of the edge eigenstates, characterized by Eq.~(\ref{eq:basis}) in the main text, when orbital motion of electrons is included.  
The orbital motion of an electron in an external magnetic field is described by minimal coupling 
$\mathcal{H}(\textbf{k} - e\textbf{A}/(\hbar c))$, where we have used a Landau gauge $\textbf{A} = (yB_z, 0,0)$ for the electron along $x$-edge. The orbital motion produces $B_z$ dependent terms in Eq.~(\ref{eq:edge}) that go as 
$H^{{\rm orb}}_{mn} = \bra{\Phi_{m}}\mathcal{H}_0(-eyB_z/(\hbar c ), 0)\ket{\Phi_{n}}$, with $m,n\in \{1,2\}$ index the spin-orbit mixed edge eigenstates in Eq.~(\ref{eq:basis}) of the main text. Orbital motion only arises for a perpendicular field, vanishing for in-plane fields.

We note that terms that go as $\sigma_z$ commute with the first term of Eq.~(\ref{eq:edge}) of the main text, and hence cannot gap out the edge states; instead they shift the position of the gapless Dirac point in momentum space. As a result,
we focus only on the off diagonal elements $H^{{\rm orb}}_{12} =H^{{\rm orb}*}_{21}$ which can gap the spectrum. As discussed in the main text, we use the numerical edge eigenstates NE method to $H^{\rm orb}$. Comparing with Eq.~(\ref{eq:edge}) we plot the gyromagnetic coefficients $g_{xz}$ and $g_{yz}$ shown in Fig.~(\ref{f2}) of the main text. 
 
The same analysis can be performed using the approximate edge eigenstates that proceed from Eq.~(\ref{eq:unitary_matrix}). Using these approximate edge eigenstates we find
\begin{multline}
\label{eq:off_diag_element}
H^{{\rm orb}}_{12} \approx \cos\frac{\chi_1}{2}\sin\frac{\chi_1}{2} \left(\mathcal{M}_{++} - \mathcal{M}_{--}\right) \\ + \sin^2\frac{\chi_1}{2}e^{-i\chi_2}\mathcal{M}_{-+} - \cos^2\frac{\chi_1}{2}e^{i\chi_2}\mathcal{M}_{+-}
\end{multline}
where $\mathcal{M}_{ij} = \bra{\Psi_{i}}\mathcal{H}_0(-eyB_z/(\hbar c ), 0)\ket{\Psi_j}$ matrix elements composed of the initial zero modes $i,j\in s$. Writing out $\mathcal{M}_{ij}$ dependence on WTe$_2$ parameters gives 
\begin{multline}
\cos\frac{\chi_1}{2}\sin\frac{\chi_1}{2}\left(\mathcal{M}_{++} - \mathcal{M}_{--}\right) \\ \approx -\frac{e v _x \delta_z\delta_x}{\hbar c  \delta^2}\mathcal{K}^2\left[ \frac{2C_y^2(v_y^2M + \delta^2 C_y)}{v_y^2(v_y^2M - \delta^2 C_y)}  \right]B_z
\end{multline}
and 
\begin{multline}
\label{eq:term_o}
\sin^2\frac{\chi_1}{2}e^{-i\chi_2}\mathcal{M}_{-+} - \cos^2\frac{\chi_1}{2}e^{i\chi_2}\mathcal{M}_{+-} \\ \approx\frac{e v _x}{\hbar c}\mathcal{K}^2\left[ \frac{2C_y^2\delta_x\delta_z}{v_y^2\delta^2} + i \frac{C_y\delta_x}{2M^2} \right]B_z
\end{multline}
valid in the small $\eta$ and small $\Omega$ limit. Plugging into Eq.~(\ref{eq:off_diag_element}) produces an approximate $H_{12}^{{\rm orb}}$. 
Finally, comparing with $\mu_B g B/2$, the real part of $H_{12}^{{\rm orb}}$ gives $g_{xz}$ and the (negative) imaginary part yields $g_{yz}$ as 
\begin{equation}
g_{xz}  \approx  -\frac{8m_e v_x C_y^2 \delta_x\delta_z}{\hbar^2  v_y^3 M} + \mathcal{O}(\eta^2), 
\quad g_{yz} \approx  - \frac{m_e v_x v_y \delta_x}{\hbar^2  M^2}, 
\label{eq:g}
\end{equation}
As discussed in the main text, $g_{xz}$ requires both $\delta_x,\delta_z \neq0$ manifesting a {\it cooperative} effect of COM -- this yields sizable spin canting and a significant COM of several Bohr magnetons. In contrast, $g_{yz}$ only requires $\delta_x \neq 0$ but is suppressed taking on values of $g_{yz} \sim 0.02$ that is suppressed by a large $M$. Here we have used the same WTe$_2$ parameters as the main text.

\subsubsection{Spin Zeeman magnetoresponse}

\begin{figure}
\includegraphics[width=\columnwidth]{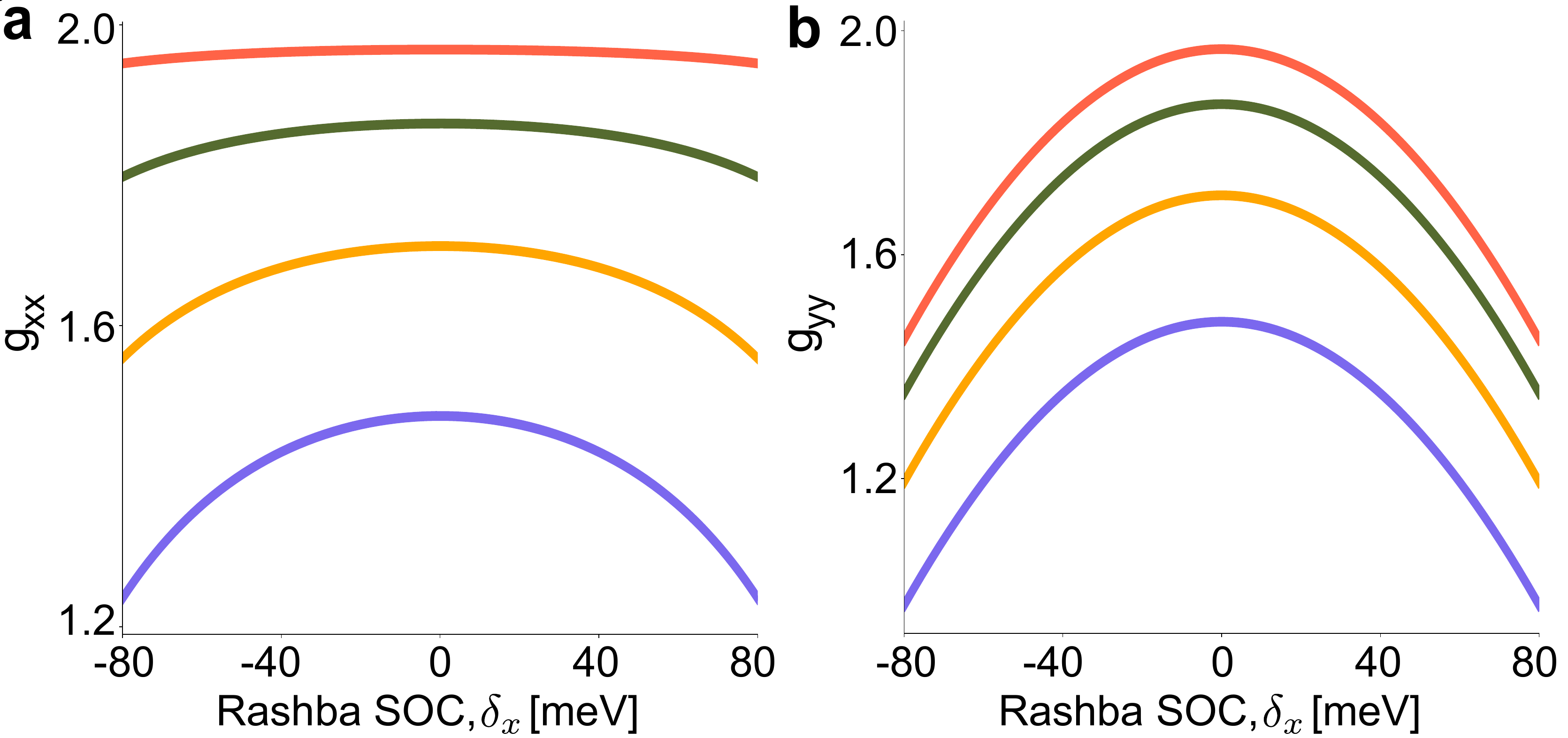}
\caption{Variation of $g_{xx}$ (panel a.) and $g_{yy}$ (panel b.) with Rashba SOC at different magnitude of Ising SOC (top to bottom: 20, 40, 60, 80 meV).}\label{g_xx+g_yy}  
\end{figure}

We now turn to the {\it spin} Zeeman interaction, wherein the magnetic field directly couples with the {\it spin moments} of the electrons. This starkly contrasts with the {\it orbital} effect of COM which we focussed on in the main text. Following the projection procedure described above, the Zeeman effect on the edge can be obtained by projecting bulk spin Zeeman interactions on the edge eigenstates as 
$[H^{\rm edge}_{Z}]_{mn} = \bra{\Phi_{m}}\mathcal{H}_Z(\textbf{B})\ket{\Phi_{n}}$, $m,n \in \{1,2\}$ where 
\begin{multline}
\label{eq:zeemanSI}
 \mathcal{H}_Z (\textbf{B})= \frac{\mu_B}{2} \left[s_z B_z\left(\mathcal{G}^+_{\perp}\tau_0 + \mathcal{G}^-_{\perp}\tau_z\right) \right. \\ \left.+ (s_x B_x+s_y B_y)\left(\mathcal{G}^+_{||}\tau_0 + \mathcal{G}^-_{||}\tau_z\right)\right]
 \end{multline}
where $\mathcal{G}^{\pm}_{\perp,||} = \left(g^e_{\perp,||} \pm g^h_{\perp,||}\right)/2$ with $g^{e,h}_{\perp,||}$ being the bulk gyromagnetic coefficients for electron/holes in response to out-of-plane/in-plane magnetic field. In our plots, we use bulk Lande g-factors of order 2, so that $\mathcal{G}^+_{||} = 2$. Other values of $g$ can be used as well, and do not qualitatively affect our results 

Using Eq.(\ref{eq:zeemanSI}) we plot the effective (in-plane) gyromagnetic coefficients $g_{xx}$ and $g_{yy}$ for the edge electrons are shown in Fig.~\ref{g_xx+g_yy}. In so doing we have used the numerical edge eigenstates using the NE method.
Rashba and Ising couplings can renormalize these coefficients so that ${\rm max}(g_{xx},g_{yy}) = \mathcal{G}^+_{||}$ when 
$\delta_x,\delta_z = 0$. In the range of Rashba and Ising SOC we consider, these produce $g_{xx}, g_{yy} \sim 1.2-2$

We note that when magnetic field is applied in the $z$-direction, the (spin) Zeeman interaction does not open a gap in the edge spectrum as discussed in Ref.~\cite{magnetoconductance, tarasenko, ronny_zeeman}. Instead, it shifts spectrum in momentum space.
This directly proceeds from $[\mathcal{H}_0(k_x,0), \mathcal{H}_Z(B_z)]=0$ as can be readily verified from Eq.~(\ref{eq:zeemanSI}). Here $\mathcal{H}_0(k_x,0)$ is the dispersing part of bulk hamiltonian that produces the gapless first term of Eq.~(\ref{eq:edge}) of the main text (i.e. the QSH gapless edgestates at $B=0$). Projecting on the edge yields $[H^{\rm edge}(\vec B=0), H^{\rm edge}_Z(B_z)] = 0$ for any choice of spinor basis.  As a result, the (spin) Zeeman interaction along the edge does not open a gap in the edge spectrum when $\vec B = B_z \hat{\vec{z}}$.

Including both
the orbital and (spin) Zeeman contributions we obtain
the variation of (full) gap size with orientation of magnetic field, see Fig. \ref{fig1} of the main text, as 
\begin{multline}
\Delta = \mu_B |\textbf{B}| \left[(g_{xx}\cos\varphi\sin\theta + g_{xz}\cos\theta)^2 \right. \\ \left.+ (g_{yy}\sin\varphi\sin\theta + g_{yz}\cos\theta)^2\right]^{1/2}.
\end{multline}
Specializing to $\varphi=0$ and noting that $g_{yz} \ll g_{xz},g_{xx}$ produces Eq.~(\ref{eq:gap}) of the main text. 

\subsection{Edge states along other edge orientations}

\begin{figure}
\includegraphics[width=\columnwidth]{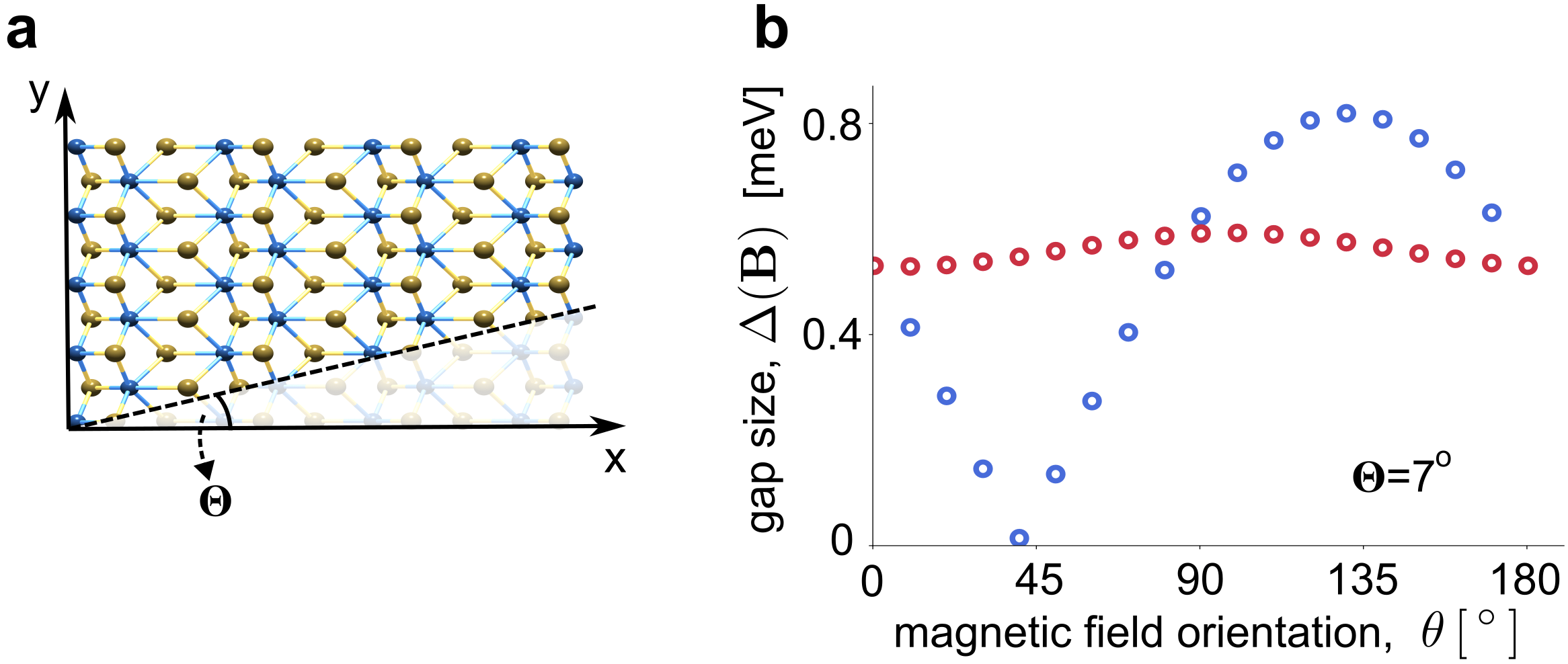}
\caption{(a) Illustration of an edge at an angle $\Theta$ from the $x$-edge so that the electron momenta  between two edges are related by Eq. (\ref{rotation_new}). (b) Gap size (calculated using ENS method, see above) for an edge oriented at $\Theta = 7^{\circ}$ from the $x$-edge. (Blue circles) Magnetic field is applied $x-z$ plane ($\phi=0$) plane; (Red circles) Magnetic field is applied in the $y-z$ plane. 
We have used $|\vec{B}| = 7$ T with $\delta_x = 40$ meV and $\delta_z = 70$ meV for Rashba and Ising SOC strength respectively. Here we have additionally chosen $\mathcal{G}^-_{\parallel} = 0.05$ as an illustration. }\label{gap_y_7_edge}
\end{figure}

While in the main text we focused on the behavior of edge states along the $x$-edge, we now consider other edge orientations. This can be done by fixing the mirror axis in absolute space (i.e. parallel to $x$-edge), and considering an edge at an angle to the $x$-edge (see Fig.~\ref{gap_y_7_edge}a). To track the electrons along this $\Theta$-edge, we use new momenta $\vec{k}'= (k_x', k_y')$, where $k_x'$ is parallel to the $\Theta$-edge and $k_y'$ is perpendicular to the $\Theta$-edge. Using these new momenta we can express the Hamiltonian in Eq.~{eq:numerical} in terms of the new momenta by  writing 
\begin{equation}
\label{rotation_new}
\begin{pmatrix}
k_x \\
k_y
\end{pmatrix} = \begin{pmatrix}
\cos\Theta & \sin\Theta \\
-\sin\Theta & \cos\Theta
\end{pmatrix}\begin{pmatrix}
k'_x \\
k'_y
\end{pmatrix}
\end{equation}
and $\Theta$ is the angle between orientation of new edge and $x$-edge. We calculate the edge state spectrum using the ENS scheme as described above with the same Dirichilet boundary conditions applied instead now on the $\Theta$-edge. 
The corresponding gap in the edge spectrum in presence of magnetic field can be similarly obtained directly ENS (see Fig. \ref{gap_y_7_edge}), where we substitute $k'_y \rightarrow -i\partial_{y'}$ and $k'_x \rightarrow (k'_x - ey'B/(\hbar c))$ as before. As a demonstration we show the edge gap opening for $\Theta = 7^\circ$ demonstrating that the large gap opening from an orbital contribution persists in Fig.~\ref{gap_y_7_edge}.

\begin{figure}
\includegraphics[width = \columnwidth]{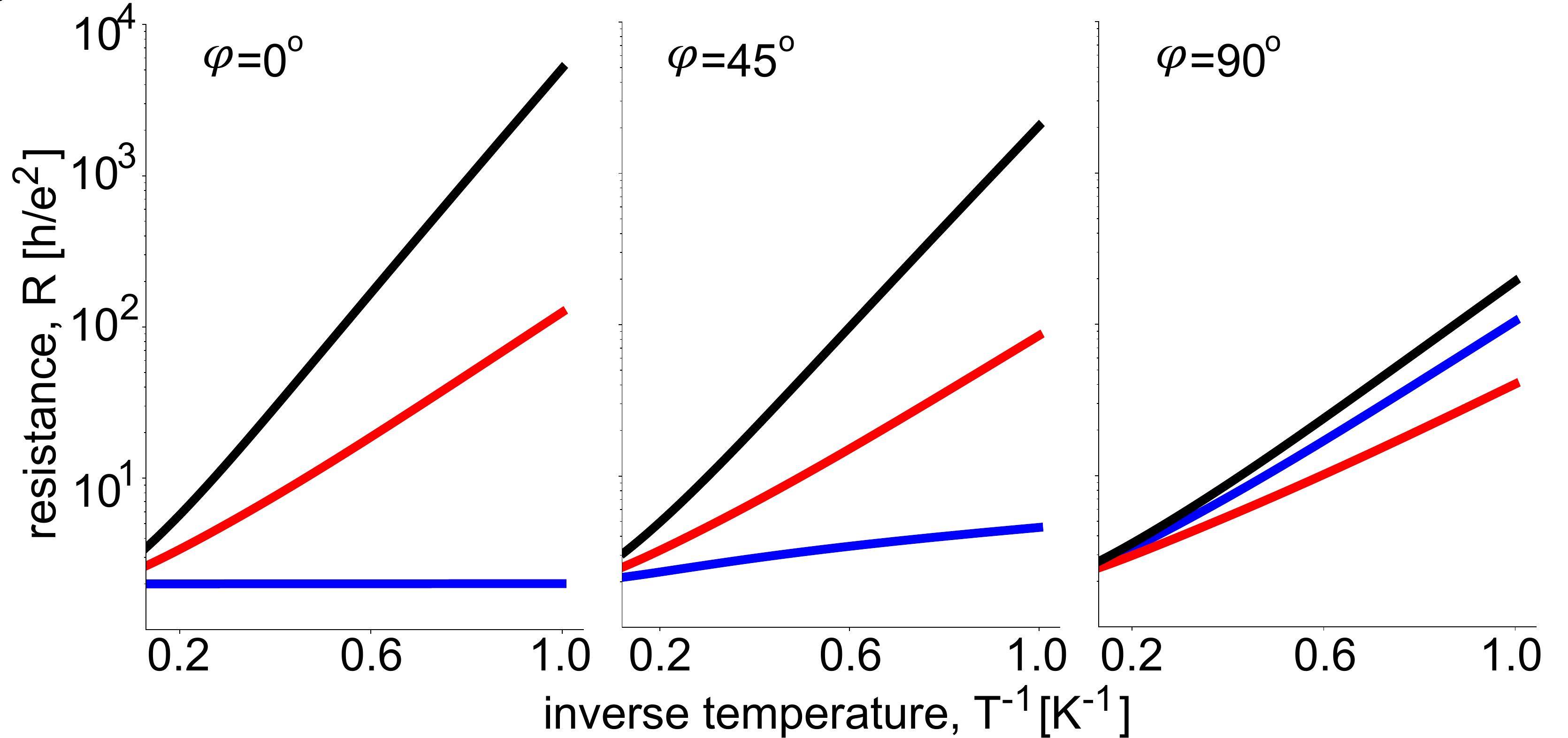}
\caption{Temperature variation of resistance at different orientations of magnetic field shown at $\theta =50 \text{ (blue)}, 95 \text{ (red)}, 140 \text{ (black)}$  for $\varphi = 0, 45^{\circ} \text{ and } 90^{\circ}$.  Anisotropic trends are seen for $\varphi=0$ which subside as the azimuthal is changed from $0$ to $90^{\circ}$. Here, $|\vec{B}|=7$ T with chemical potential fixed at $\mu = 0.3$ meV.} \label{r_0_45_90}
\end{figure}

\subsection{Edge resistance at various $\varphi$} 
In this section, we investigate the temperature dependence of resistance for various azimuthal angles and we find anisotropic (magnetoresistance) temperature dependence of resistance at 
is most pronounced for $\varphi=0$.  As the gap almost closes at $\vec{B}_{*}$,  for any finite chemical potential, there are free carriers in the conduction band and we see a minimum resistance at $\theta = \theta_c$, that is insensitive to temperature (panel a in Fig.~\ref{r_0_45_90}, blue curve). 
In contrast, as we change the azimuthal angle from 0 to 90$^{\circ}$, the gap size at $\theta=\theta_c(90+\theta_c)$ increases (decreases) significantly, and the minima (maxima) in gap shifts towards 90$^{\circ}$ ($0^{\circ}$). This reduces the anisotropy in the temperature variation in resistance, which is most evident at $\varphi=0$, see Fig.~\ref{r_0_45_90}. In plotting the resistance in Fig.~\ref{r_0_45_90} we have used the same parameters as the main text.

\end{document}